# Experimental Investigation of Skew Wind Effects on Vortex-Induced Vibration of Typical Bridge Decks

Guangzhong Gao[a,*], Pengwei Zhang[a], Wenkai Du[a], Yonghui Xie[a], Pengjie Ren[b], Xiaofeng Xue[a]

a. Highway College, Chang'an University, Xi'an 710064, China;

b. China Design Group Co., Ltd., Nanjing 210014

**Abstract**: This study explores the skew wind effects on vortex-induced vibration (VIV) of two typical bridge decks—a closed-box girder and a twin-edge girder—through spring-suspended oblique section model tests. Experiments were conducted at various wind yaw angles ($\beta_0$= 0°, 5°, 10°, 15°, 20°, 25°) and angles of attack ($\alpha_0$ = -3°, 0°, 3°). Results indicate that VIV amplitudes and lock-in ranges exhibit a clear variation with yaw angles, rendering the Independence Principle (IP) unsuitable for these configurations. A heave-torsion coupling phenomenon was observed in both heaving and torsional VIV, attributed to the eccentricity of the center of gravity due to asymmetric end segments. A novel numerical algorithm was introduced to correct peak VIV amplitudes for variations in structural mass-damping parameters across yaw angles. The most unfavorable VIV responses occurred under skew wind conditions, with maximum amplitudes increasing by approximately 20.1% for heaving VIV and 179.8% for torsional VIV in the closed-box girder, and by 3.9% for heaving VIV in the twin-edge girder, relative to normal wind conditions.

## 1. Introduction

Vortex-induced vibration (VIV) is a critical aerodynamic instability where periodic vortices alternately shed from a bluff body, exciting the structure into resonance with large-amplitude limit cycle oscillations. VIV is a key design consideration for various engineering structures, including flexible bridge components, chimneys, offshore risers, and transmission lines. Several bridges, such as the Tokyo-Bay Sea-Crossing Bridge [1], Xihoumen Bridge [2], Humen Bridge [3], Hong Kong-Zhuhai-Macao Bridge [4], and Volgograd River Bridge [5], have experienced large-amplitude VIV. Although VIV typically does not cause structural collapse, sustained oscillations can disrupt traffic, threaten driving comfort and safety, and induce fatigue damage at member connections [6].

Due to its frequent occurrence at low wind speeds, VIV has attracted wide attention in wind engineering [6]. Wind tunnel experiments and numerical simulations have extensively explored influential factors, including freestream turbulence [7], Reynolds number [8], interference effects [9], wind angles of attack [10], structural mass-damping parameters [11], aerodynamic mitigation measures [12-14], mechanical control using tuned mass dampers (TMD) [15], and analytical theories in hybrid time-frequency domains [16-17], pure time domains [18-19], and Griffin plots [20-21].

However, most existing studies focus on normal flow conditions, where the flow vector is perpendicular to the bridge deck's longitudinal axis (i.e., wind yaw angle $\beta_0$ = 0°). The effect of wind yaw angle on VIV responses, particularly for bridge deck cross-sections, remains underexplored. In practice, bridges are frequently exposed to skew wind. Traditionally, the "Independence Principle" (IP), or "cosine rule" [22-23], is applied to assess skew wind effects. The IP postulates that the axial component of yawed flow ($U\sin\beta_0$) does not influence vortex-induced forces, and VIV responses depend solely on the normal component ($U\cos\beta_0$). If valid, the IP implies that the VIV lock-in range, maximum amplitude, and Strouhal number under various yaw angles should match those under normal flow when using the normal component $U\cos\beta_0$ to calculate reduced velocity [24-26].

The IP's validity has been evaluated through experiments on flexible and rigid circular cylinders. Xu et al. [24] found that for flexible cylinders, the IP holds for multi-mode VIV responses at $\beta_0 \leq 30°$, but discrepancies emerge at 45°–60°. Bourguet et al. [25] used direct numerical simulations of a flexible circular cylinder at $\beta_0 = 60°$, concluding that the IP applies to high-tension cylinders with minimal in-line bending, but not to lower-tension configurations with significant in-line bending. Franzini et al. [26] observed in water tunnel experiments that the IP is valid for yaw angles up to 20° for one- and two-degree-of-freedom systems, with reduced maximum amplitudes beyond 20°. Wang et al. [27] numerically studied a rectangular 4:1 cylinder, suggesting the IP holds for $\beta_0 \leq 30°$, yet subsequent wind tunnel tests showed decreasing VIV amplitudes with increasing yaw angles, invalidating the IP [28]. Xu et al. [29] tested a bluff box-girder with a side ratio of 3.8, noting that VIV responses peak under normal wind, with amplitudes decreasing as yaw angles increase.

For flutter instability, the IP was historically used to simplify skew wind analysis, as Scanlan [30] noted: "a given skew wind velocity will be critical for flutter if its cosine component normal to the bridge deck equals the critical bridge normal-wind velocity." Zhu et al. [31-33] extensively tested this hypothesis via wind tunnel experiments, finding the IP invalid for common bridge decks. The lowest flutter onset velocity often occurs at yaw angles between 5° and 20°, with critical flutter velocity varying non-monotonically. For wind angles of attack between -3° and 3°, flutter onset velocity can drop by 6% for flat box-girder decks, 11% for flat π-shaped sections, and 16% for truss-type decks with narrow central gaps compared to normal wind conditions. Aerodynamic measures, such as lower central stabilizing barriers on truss-type girders, can reduce flutter onset velocity by approximately 22% under skew winds. Similar trends appear in buffeting responses, where the most unfavourable conditions often arise under yawed wind [34-35].

It is clear from the above literature review that the IP is valid only for circular cylinders at small yaw angles, becoming inapplicable at larger angles or for non-circular sections. While some studies suggest VIV amplitudes decrease with increasing yaw angles, this is not conclusive for complex bridge deck sections. Given that yawed winds significantly affect vortex strength and wake patterns [36-37], the nonlinearity introduced by bridge deck shapes likely invalidates the IP. Moreover, experimental evidence on flutter suggests that unfavourable flutter instability may occur under yawed winds rather than normal winds. Thus, investigating VIV responses under skew winds is essential to determine whether peak amplitudes occur under normal or yawed conditions.

This study examines skew wind effects on VIV responses of two representative bridge decks—closed-box and twin-edge girder—via oblique section model tests. Key objectives include assessing the IP's validity for heaving and torsional VIV and exploring a novel heave-torsion coupling phenomenon under skew winds, previously unreported in the literature.

The paper is structured as follows: Section 2 describes the experimental setup for VIV tests, focusing on the oblique section model configuration and wind yaw angle adjustments. Section 3 presents VIV responses under various wind angles of attack and yaw angles, with detailed analysis of the observed heave-torsion coupling. Section 4 introduces a numerical algorithm to adjust VIV peak amplitudes across yaw angles and examines their variation with skew angles. Section 5 summarizes the main findings.

## 2 Wind tunnel experiments

### 2.1 Experimental setup

The wind tunnel tests on vortex-induced vibration (VIV) under skew wind conditions were conducted in the CA-1 wind tunnel at Chang'an University. The CA-1 wind tunnel is a closed-circuit atmospheric boundary layer tunnel with a test section 3m in width, 2.5 in height, and 15m in length. The background turbulence intensity $I_u$ is less than 1%, and the wind speed can be continuously adjusted across a range of 0.5 to 53m/s.

The experimental configuration, illustrated in Figure 1, features oblique section models elastically supported by eight pretensioned helical springs, enabling heaving and torsional vibrations. As depicted in Figure 2, each oblique section model comprises five components: one middle segment with a rectangular outline, two transverse rigid arms for suspending the helical springs, and two trapezoidal end segments to ensure that the model's ends remain parallel to the oncoming flow. These five segments are rigidly assembled end-to-end using crews to form a cohesive section model. For each tested skew angle, the trapezoidal end segments were replaced to align the model's ends with the incident flow direction, while their lengths were held constant across different yaw angles to ensure consistent mass properties.

Figure 2 also defines the wind yaw angle $β_0$ and wind angle of attack $α_0$. The wind yaw angle $β_0$ is the relative angle between the direction of the oncoming flow and the normal direction of the longitudinal axis of the middle segment, and positive when the model rotates clockwise. The wind angle of attack $α_0$ is defined as the pitch angle of the middle segment's cross-section relative to the horizontal plane, considered positive if upstream nose up.

The oblique section model was suspended between the steel ceiling and floor of the wind tunnel via magnetic bases. To adjust the wind yaw angle $β_0$, the magnetic bases were first deactivated, allowing manual rotation of the section model to the target orientation. Once the model was positioned, the magnetic bases were reactivated to fix both the model and springs in place. No end plates were attached to the mode, permitting the development of axial flow under skew wind conditions. The adjustment of wind attack angle $α_0$ was achieved by extending or shortening the turnbuckles.

Dynamic displacement signal of the oblique section was measured using two laser displacement sensors positioned at the downstream suspending arm. No sensors were placed on the upstream side to avoid potential interference with the flow field.

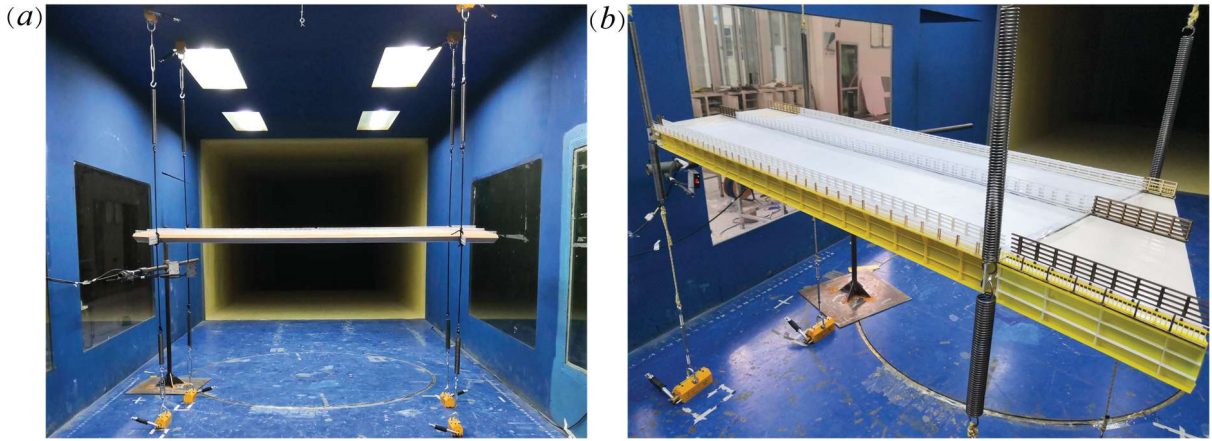

Figure 1. Elastically supported oblique section models under skew wind. (a) the closed box bridge deck, (b) twin-edge girder bridge deck.

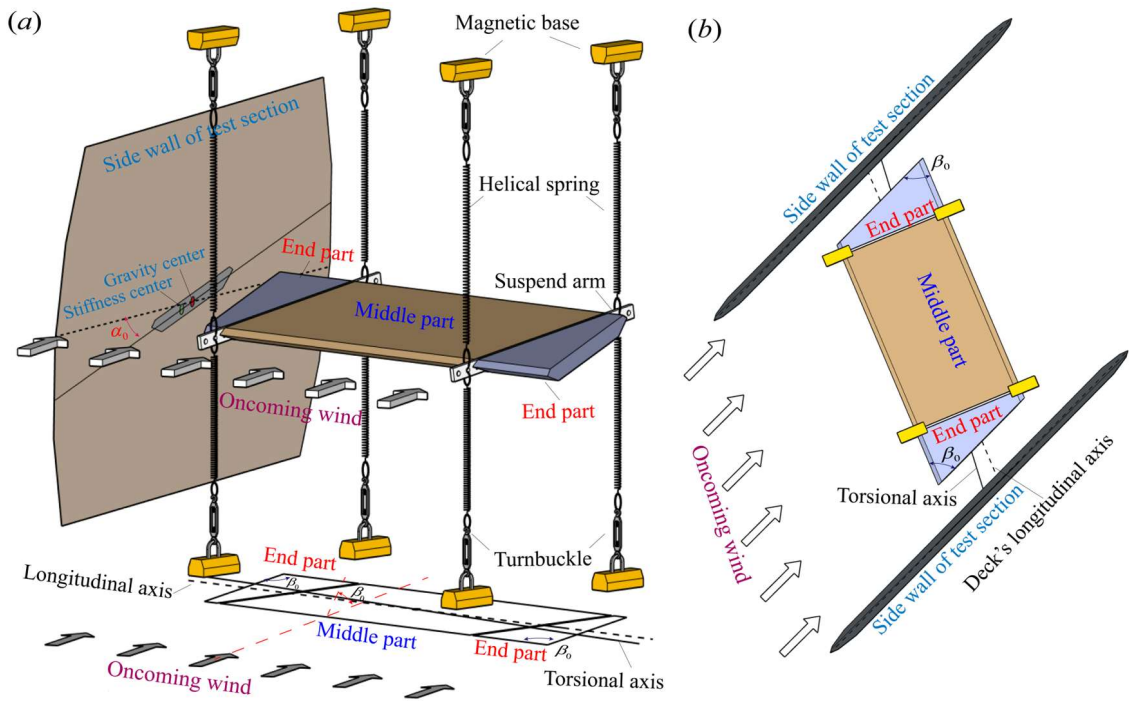

Figure 2. Schematic of the experimental setup for the free vibration under skew wind: (a) oblique view, (b) top view.

## 2.2 Dynamic parameters

The Humen Bridge and Wangguan Bridge, with their closed-box and twin-edge girder decks, respectively, provide the engineering background for this study. Section model was designed with geometric length scale $\lambda_L$ of 1:60 for the closed box bridge deck and 1:50 for the twin-edge girder section. The scaled cross-sections are illustrated in Figure 3 and Figure 4. The section models were fabricated with lightweight yet stiff materials, such as wooden plates and foam to simulate the aerodynamic configuration. Two metal tubes were installed inside each model to ensure overall rigidity. Crash barriers, handrails were engraved from plastic/wooden plates and assembled.

Table 1 lists the main geometric and dynamic parameters of the oblique section models. The aspect ratios were designed to be sufficiently large, specifically 3.66 for the closed-box girder model, and 3.08 for the twin-edge girder model, to minimize end effects. The nominal Reynolds number $Re_0 = UD/\nu$ of tested VIV lock-in regions is calculated using the free-stream velocity $U$, and effective Reynolds number $Re_\perp = UD\cos\beta_0/\nu$ under yawed wind can be further determined considering wind yaw angle $\beta_0$.

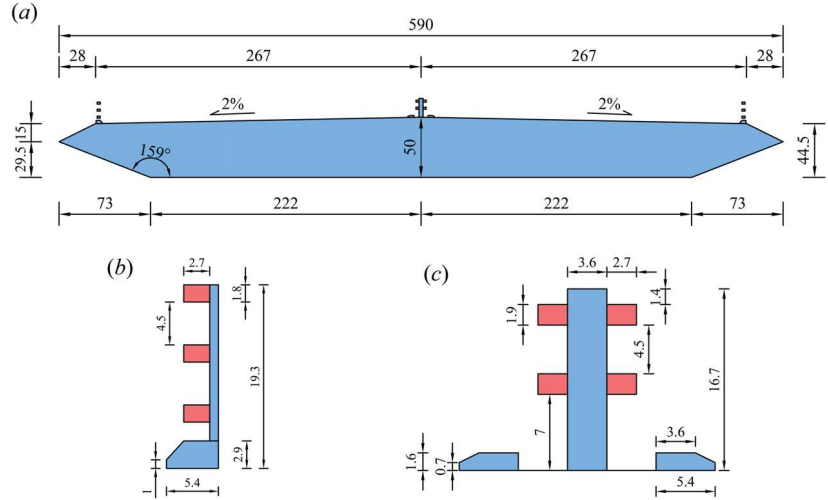

Figure 3. Cross-sectional dimensions of the closed-box section (Unit: mm)

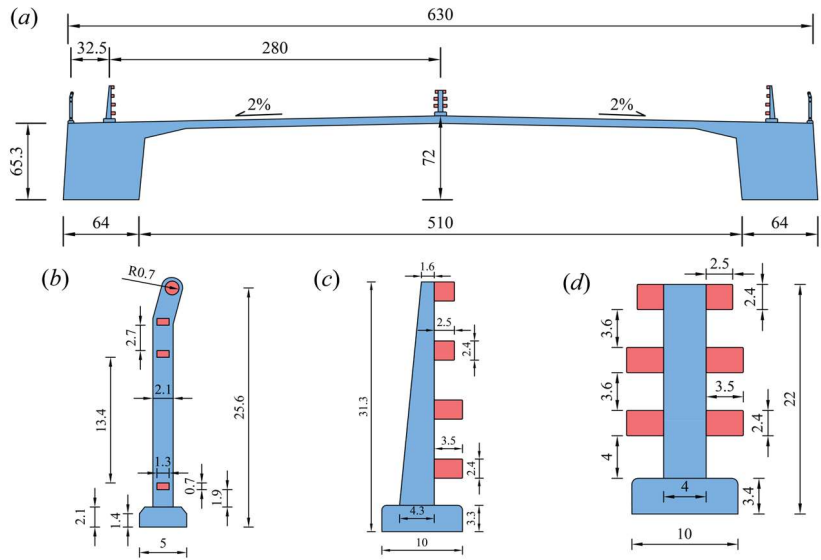

Figure 4. Cross-sectional dimensions of the twin-edge girder section (Unit: mm)

Table 1. Main structural parameters of the oblique section models.

| Parameter | Closed-box girder model | Twin-edge girder model |
|---|---|---|
| Section width $B$ (m) | 0.590 | 0.630 |
| Section depth $D$ (m) | 0.050 | 0.072 |
| Model length $L$ (m) | 2.160 | 1.937 |
| Effective mass $M$ (kg/m) | 6.771 | 24.663 |

| Effective mass moment of inertia $J_m$ (kg·m²/m) | 0.431 | 0.603 |
| Heaving frequency $f_h$ (Hz) | 4.0 | 3.7 |
| Torsional frequency $f_t$ (Hz) | 6.7 | 5.9 |
| Length scaling $\lambda_L$ | 1:60 | 1:50 |
| Velocity Scaling $\lambda_u$ | 1:3.24 | 1:3.2 |
| Nominal Reynolds number $Re_0 = UD/\nu$ | $1.0 \times 10^4$-$2.0 \times 10^4$ | $1.4 \times 10^4$-$3.0 \times 10^4$ |

Free vibration tests were conducted on the two bridge decks models in a smooth flow field. Each model evaluated six wind yaw angles ($\beta_0$): 0°, 5°, 10°, 15°, 20° and 25°, and three wind angles of attack ($\alpha_0$): -3°, 0° and 3°. For each yaw angle, the two end segments were replaced to align the model ends with the incident flow. Consequently, 12 sets of end segments were manufactured for the two models prior to testing.

For each test configuration, the vibration frequencies and heave/torsion damping ratios were determined from the free decay responses in still air. The heaving and torsional frequencies exhibited minimal variation across configurations and were treated as constants, as listed in Table 1. The structural damping ratios were depicted in Figure 5. These damping ratios vary noticeably across test configurations due to adjustments in wind angles of attack and yaw angles, which involve modifying turnbuckles and replacing end segments. As a result, the structural damping mechanisms differ between test cases.

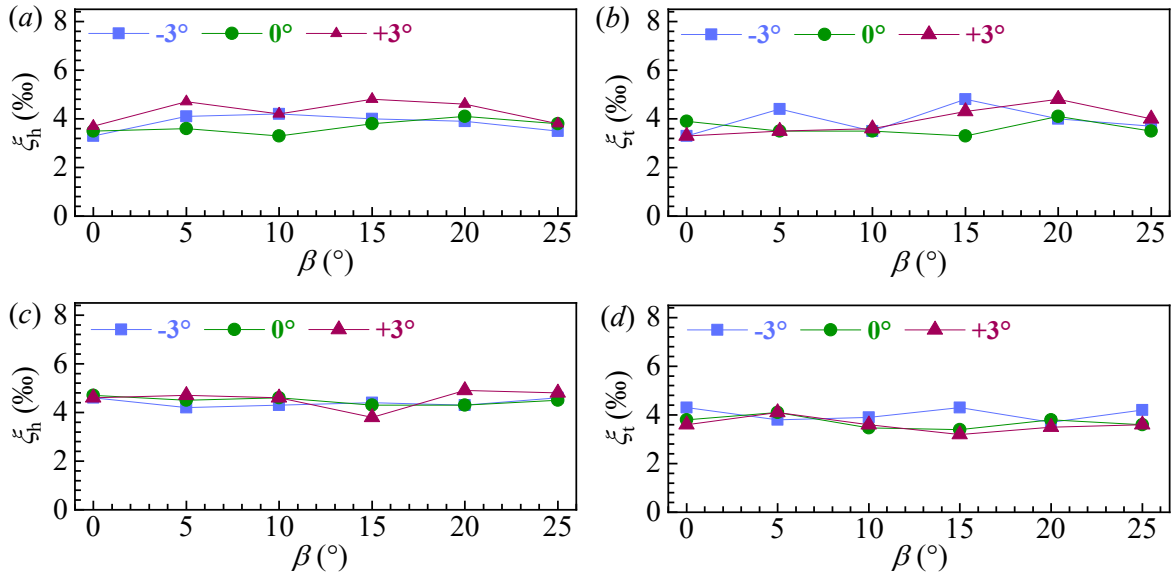

Figure 5. Structural damping ratios under various yaw angles $\beta_0$ and wind angles of attack $\alpha_0$. (a) Heaving and (b) torsional damping ratios for the closed-box girder model; (c) Heaving and (b) torsional damping ratios for the twin-edge girder model.

## 3. Experimental results

### 3.1 Skew wind effect on VIV lock-in range

The closed-box and twin-edge girder bridge deck models exhibited significant vortex-induced vibration (VIV) across various wind yaw angles ($\beta_0$). For the closed-box bridge deck, both heaving and torsional VIV were observed within separate ranges of reduced wind velocity. The peak VIV amplitudes were slightly lower than those reported by Ge et al. [3] as the present study excluded the unfavorable contribution of maintenance tracks at the girder's bottom.

Figure 6 illustrates the heaving and torsional VIV lock-in regions for the closed-box girder model at a wind angle of attack ($\alpha_0$) of 3°. The lock-in ranges and stable amplitudes vary significantly with changes in wind yaw angles. The independence principle (IP) is not applicable herein because it predicts the lock-in ranges under yawed wind be identical to the normal flow.

Another critical engineering concern is whether the peak VIV amplitude occurs under normal wind conditions ($\beta_0 = 0°$), as this determines whether conventional VIV testing under normal wind conditions yields conservative or non-conservative estimates of peak responses. As shown in Figure 6, the peak amplitude for heaving VIV is maximized at a zero yaw angle, whereas torsional VIV amplitudes increase significantly with larger yaw angles. However, this conclusion requires caution, as peak VIV amplitudes are sensitive to structural damping ratios (see Figure 5). Variations in damping ratios across yaw angles must be accounted for, as will be addressed in Section 3.

For the twin-edge girder model, heaving and torsional VIV occurred within overlapping reduced velocity ranges. At wind angles of attack of 3° and 0°, heaving VIV suppressed torsional VIV, resulting in dominant heaving VIV, as shown in Figure 7. Conversely, at a wind angle of attack of -3°, significant interactions between heaving and torsional VIV were observed, as depicted in Figure 8. Figure 7 indicates that the heaving VIV lock-in range and stable amplitudes for the twin-edge girder model varied with wind yaw angles. As with the closed-box model, comparisons of peak VIV amplitudes require adjustments for variations in structural damping ratios, which will be discussed in Section 3.

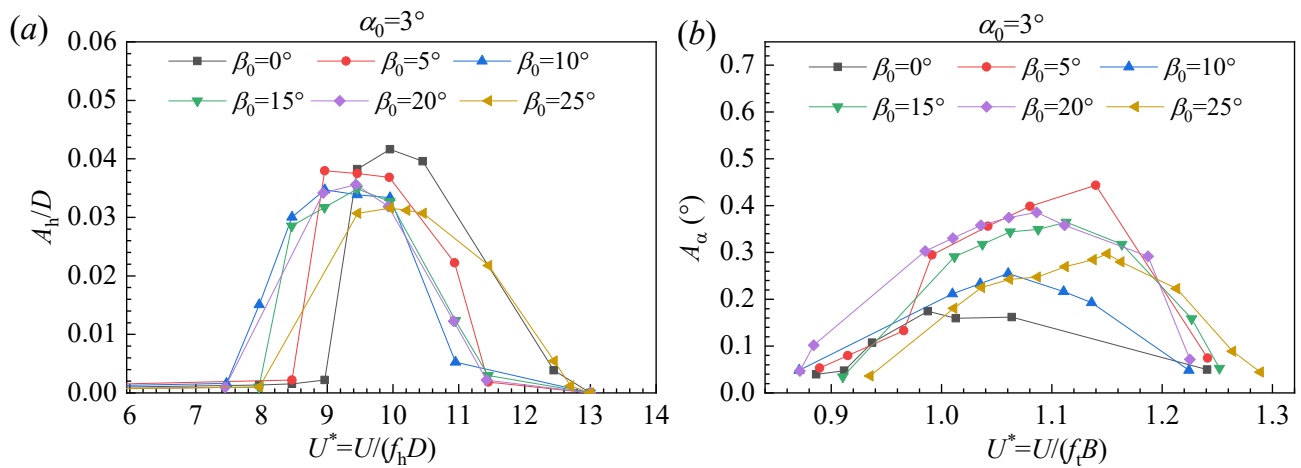

Figure 6. VIV lock-in regions of the closed-box girder model: (a) Heaving VIV, (b) Torsional VIV.

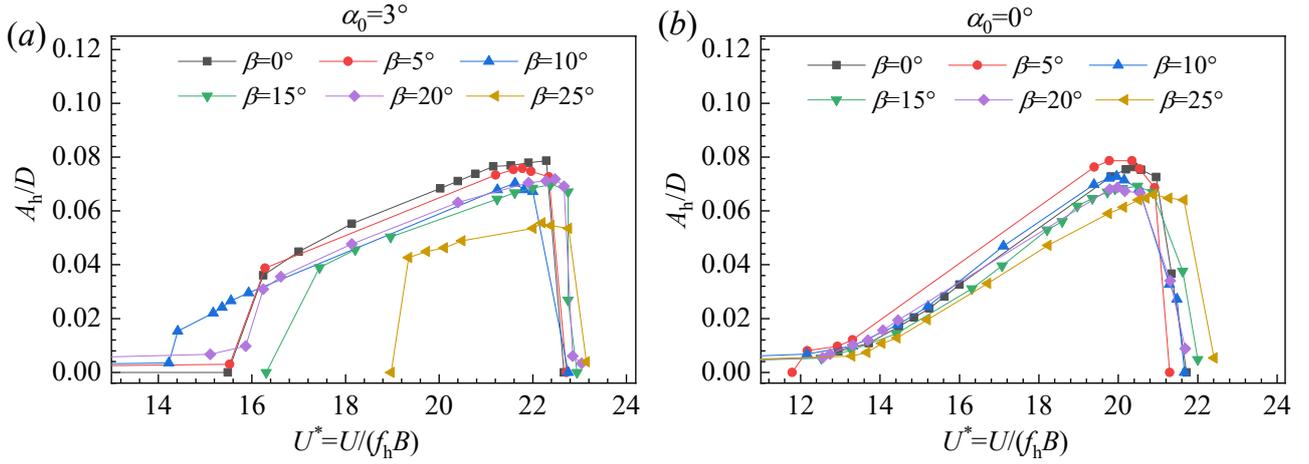

Figure 7. VIV lock-in regions for the twin-edge girder section: (a) Heaving VIV at a wind angle of attack of 3°, (b) Heaving VIV at a wind angle of attack of 0°.

Complex interactions between heaving and torsional VIV were observed for the twin-edge girder model at a wind angle of attack of -3°. As shown in Figure 8a, at a zero wind yaw angle, the model exhibited either torsional VIV or heaving VIV. Starting from rest, heaving motion initially increased alongside torsional displacement due to coexisting instabilities. However, as torsional amplitude grew, heaving motion was gradually suppressed to minimal levels upon reaching a stable torsional amplitude. This interaction was highly sensitive to initial conditions, resulting in varied displacement envelopes across different grow-to-resonance responses (labeled *A–D* in Figure 8a). For responses *C* and *D*, instability manifested in heaving motion rather than torsional motion. In summary, within the overlapping VIV range at zero yaw angle, the model exhibited either heaving or torsional VIV depending on initial conditions, with one instability suppressing the other.

Under non-zero yaw angles, the interaction phenomenon shifted. As shown in Figure 8b and 8c, increasing wind yaw angles led to dominant heaving VIV coupled with noticeable torsional motion. At a wind yaw angle of 10°, the interaction remained sensitive to initial conditions, with non-unique stable heaving amplitudes across three grow-to-resonance responses. Larger stable heaving amplitudes corresponded to smaller torsional amplitudes. At a wind yaw angle of 25°, displacement envelopes still varied, but the stable amplitudes of both heaving and torsional motion became more consistent.

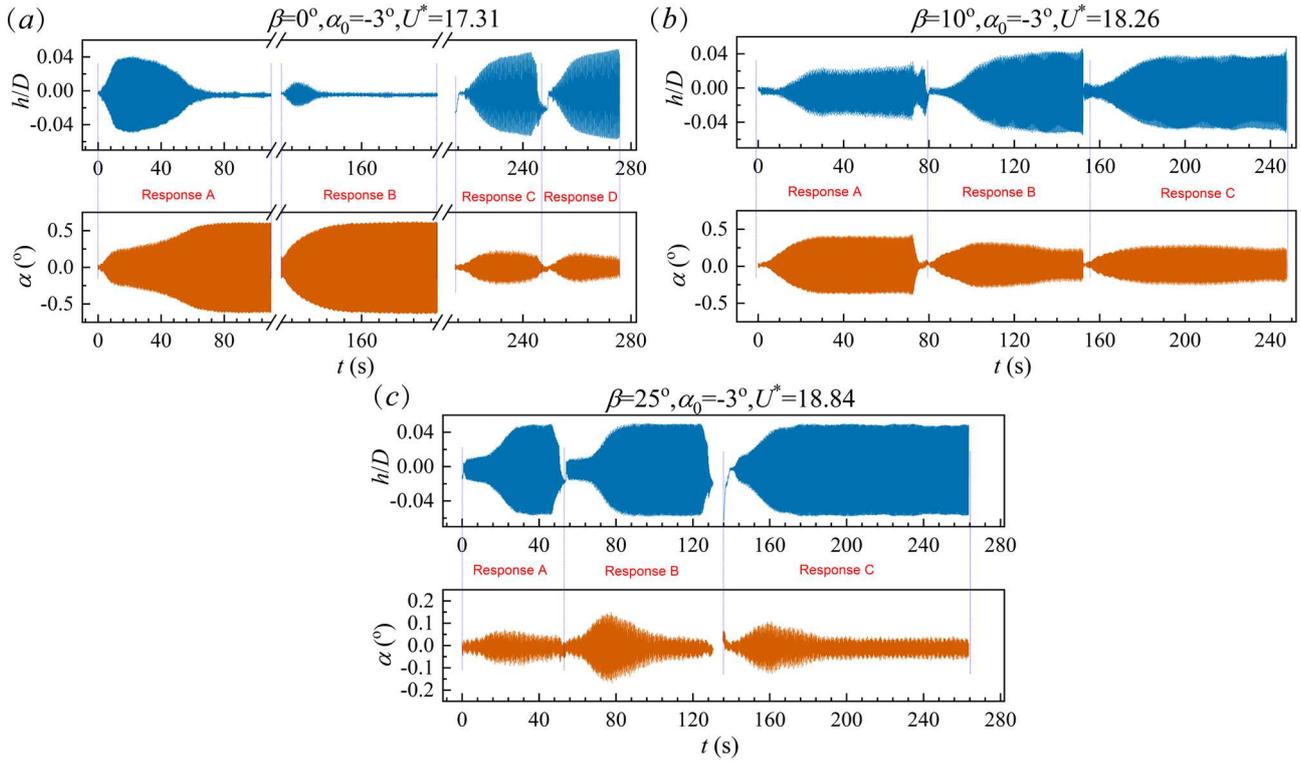

Figure 8. Interaction between heaving and torsional VIV for the twin-edge girder section at -3° wind angle of attack.

## 3.2 Heave-torsion coupled VIV under skew winds

A notable phenomenon of heave-torsion coupled VIV was observed in the oblique section model tests, which is not previously reported in wind tunnel experiments under skew wind conditions.

Figure 9 presents the displacement signals recorded by two laser displacement sensors positioned beneath the downstream suspending arm, equally spaced around the longitudinal axis of the bridge deck. As shown in Figure 9a, under normal wind conditions ($β_0=0°$), the displacement signals at the upstream and downstream points exhibit equal amplitudes and opposite phases during torsional VIV, consistent with the conventional understanding that VIV occurs predominantly in either torsional or heaving mode with minimal heave-torsion coupling. In contrast, Figure 9b shows that at a wind yaw angle, the amplitudes at the upstream and downstream points differ, while the phases remain opposite, indicating a downstream shift of the torsional axis and oscillations in a coupled mode with fixed phase differences. These displacement signals suggest that the torsional VIV is accompanied by noticeable heaving motion.

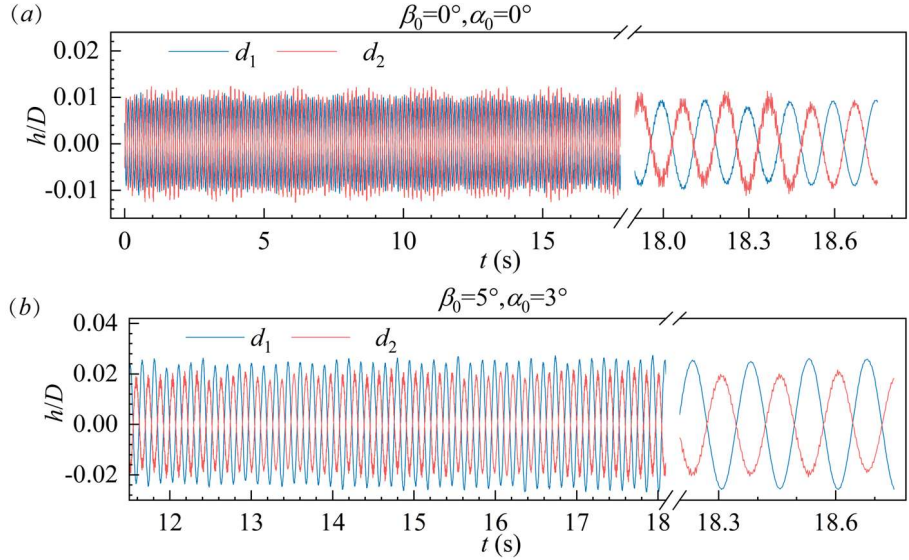

Figure 9. Time histories of displacement signals measured at the suspending arm for the closed-box girder model: (a) Torsional VIV under normal wind conditions ($\beta_0 = 0°$), (b) Torsional VIV under skew wind conditions ($\beta_0 = 5°$). The signals $d_1$, $d_2$ represent measurements at the upstream and downstream points, respectively, equally spaced around the longitudinal axis of the bridge deck.

The heave-torsion coupled VIV is further confirmed in Figures 10 and 11, which display time histories of heaving and torsional displacements during peak-amplitude stages, along with their amplitude spectra. The following observations are derived from these figures:

(1) The heave-torsion coupling effect is more pronounced in torsional VIV than in heaving VIV.

(2) No phase difference exists between torsional and heaving displacements during torsional VIV. In contrast, during heaving VIV, the torsional displacement lags behind the heaving displacement.

(3) Heave-torsion coupling occurs in both heaving and torsional VIV. For heaving VIV, the coupled torsional displacement oscillates at the heaving frequency and its higher harmonics, although the torsional frequency component is also excited, as evidenced in Figures 10b and 11b. For torsional VIV, the coupled heaving displacement oscillates primarily at the torsional frequency. Thus, the heave-torsion coupling in oblique section models manifests as a single-mode instability.

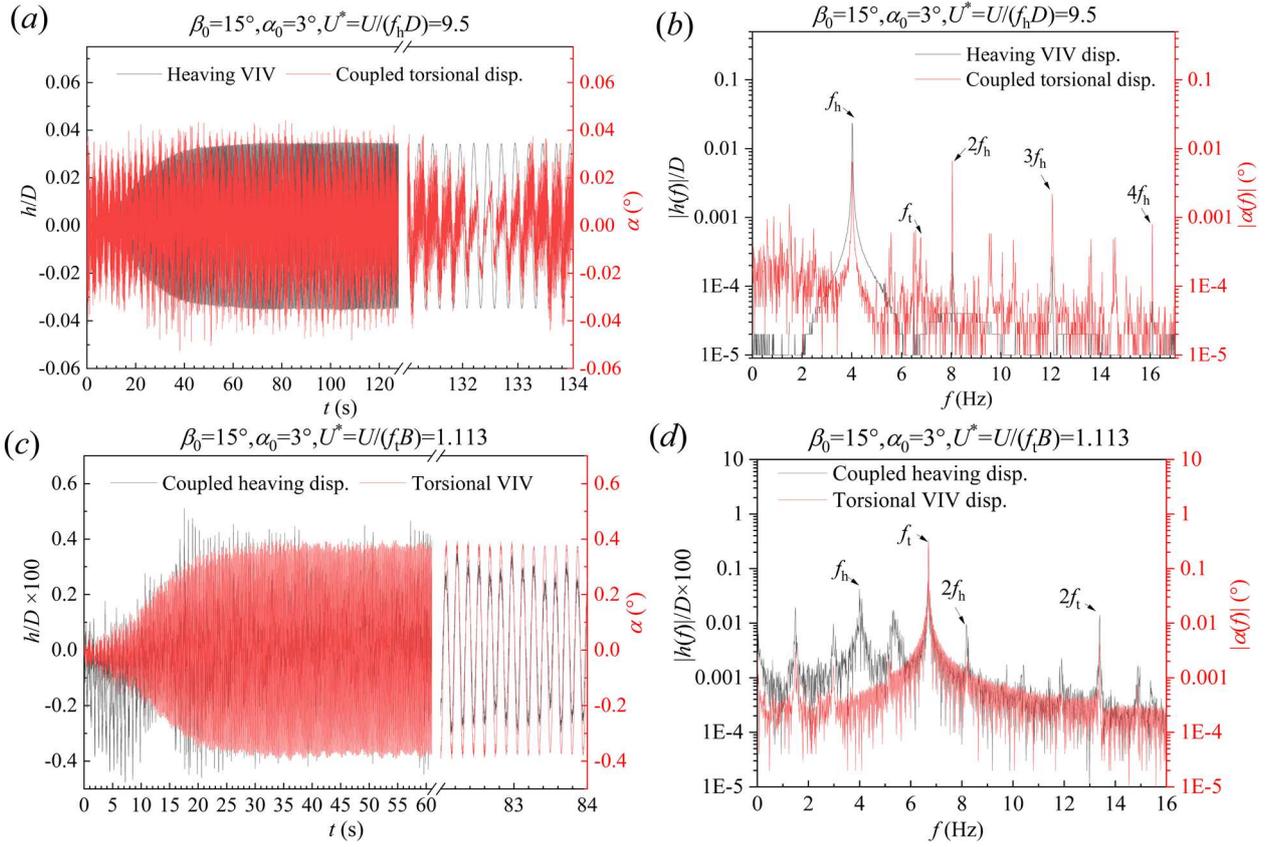

Figure 10. Heave-torsion coupled vibration for the closed-box girder model during: (a) Heaving VIV, (b) Torsional VIV.

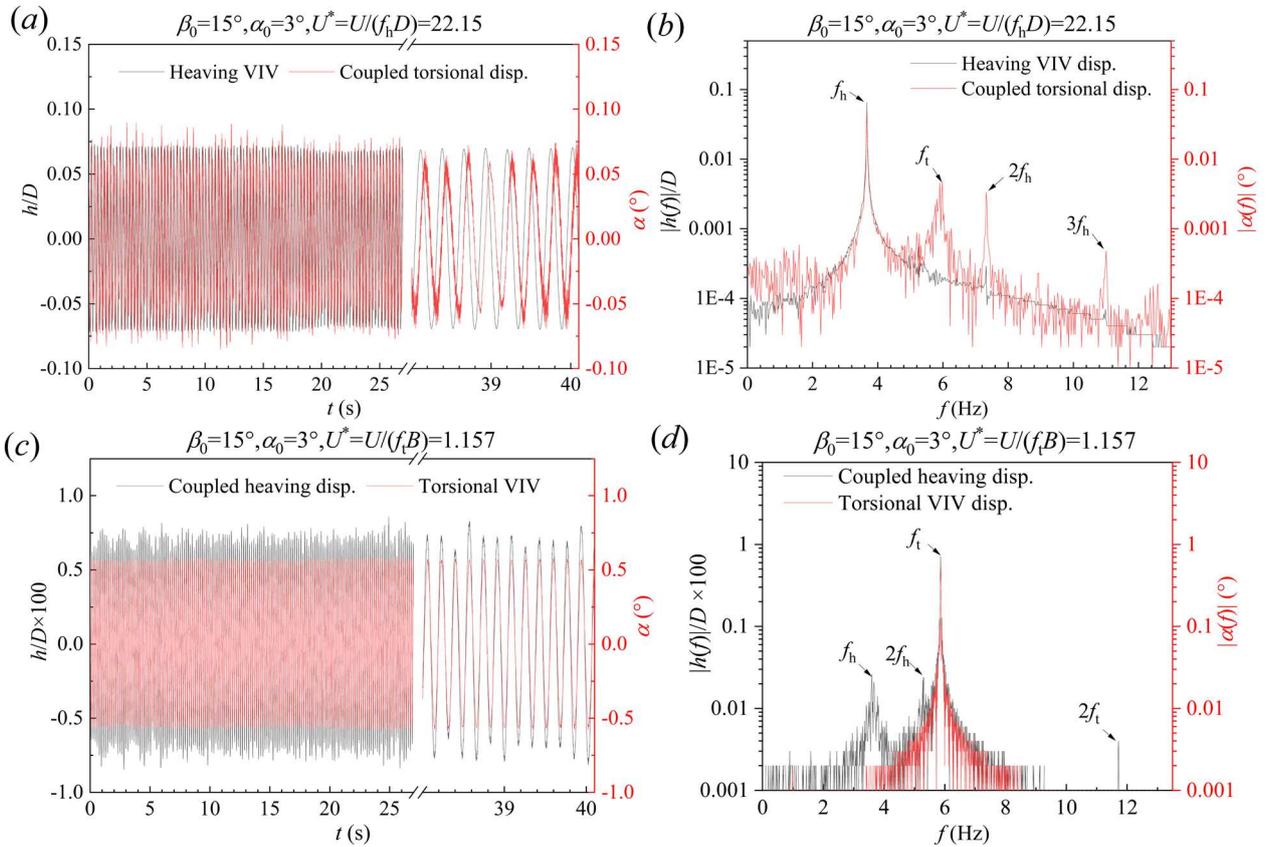

Figure 11. Heave-torsion coupled vibration for the twin-edge girder model during: (a) Heaving VIV, (b)

Torsional VIV.

Figure 12 presents phase diagrams of heave–torsion coupled oscillations for the spring-suspended section model. The horizontal axis represents heave displacement and the vertical axis represents torsional displacement. The closed loops correspond to the limit cycle stage of oscillation. Their geometry provides physical insight into the coupled behaviors:

(1) Arrows on the trajectories indicate the phase relationship and whether the torsion leads or lags the heave motion.

(2) The slope of the major axis reflects the mean coupling ratio between heave and torsion. In extreme case where the enclosed area vanishes, the trajectory collapses to a straight line, and the slope corresponds directly to the ratio of torsion-to-heave motion.

(3) A finite enclosed area indicates a phase difference between heave and torsion, which is related with aerodynamic energy exchange. A zero-area trajectory (straight line) corresponds to purely in-phase or out-of-phase motion without energy transfer.

(4) The aspect ratio of the ellipse indicates which degree of freedom dominates the motion. Elongated loops aligned with the heave axis indicates heave-dominated responses, and vice versa.

The phase diagrams of the coupled VIV cases in Figure 11 are shown in Figure 12. For comparison, it also includes the phase diagram for conventional torsional VIV [38], which occurs at high reduced velocities under normal wind conditions and also exhibits coupled heave-torsion motion; it is termed as "conventional coupled VIV" for brevity. The displacement signals for this conventional case are shown in Figure 13.

These phase diagrams differ significantly in trajectory shape, particularly in slope, phase difference, and enclosed area. The phase diagram for coupled heaving VIV in Figure 12a shows an almost zero slope, suggesting negligible eccentricity, as further confirmed in Figure 15a. In contrast, the phase diagram for coupled torsional VIV in Figure 12b has a positive slope, indicating that positive torsional displacement (upstream side upward motion) corresponds to positive heaving displacement (upward motion). Thus, for coupled torsional VIV in the present study, the torsional axis shifts downstream. For conventional coupled VIV, the negative slope indicates an upstream shift of the torsional axis.

Another key difference is the phase relationship between heaving and torsional displacement. For coupled heaving VIV, torsional displacement lags behind heaving displacement, as seen in Figure 12a and Figure 11a. For coupled torsional VIV, torsional and heaving motions are nearly synchronized with a small phase difference, though slight trajectory meandering is present in Figure 12b. In conventional coupled VIV, torsional motion also lags behind heaving motion.

Additionally, the phase diagram for coupled torsional VIV has a negligible enclosed area, indicating synchronized heaving and torsional motions characteristic of a real-mode instability. In contrast, the phase diagram for conventional coupled VIV has a significant enclosed area. Furthermore, in Figure 13, the phase difference between upstream and downstream displacement signals deviates from 180°, suggesting a complex-mode instability for conventional coupled VIV. While the phase diagram for coupled heaving VIV also suggests a complex-mode instability, the small torsional amplitude and negligible axis eccentricity allow it to be approximated as conventional pure heaving VIV.

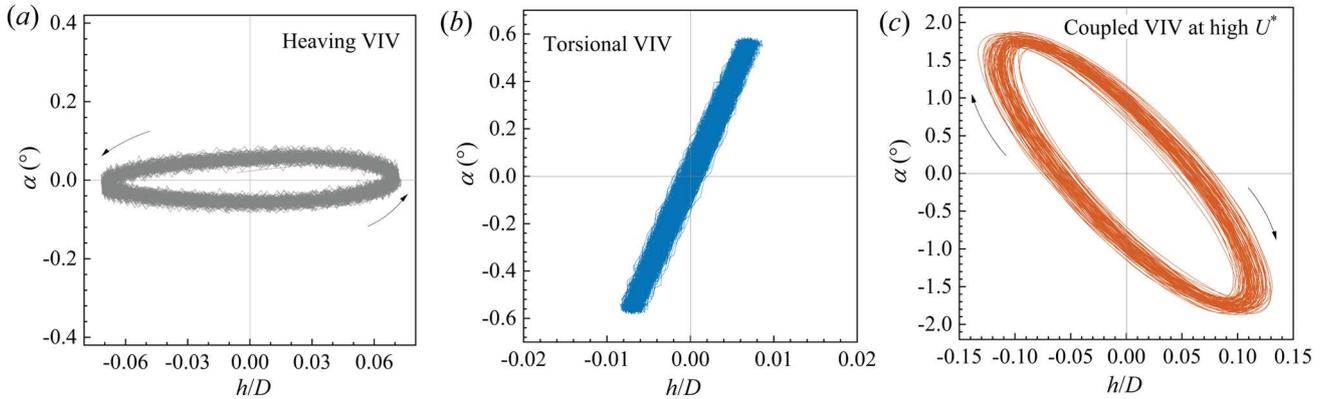

Figure 12. Phase diagrams of heaving versus torsional displacement during stable-amplitude stages for: (a) Heaving VIV in Figure 11a, (b) Torsional VIV in Figure 11c, (c) Conventional coupled VIV occurring at high reduced velocities in Figure 13.

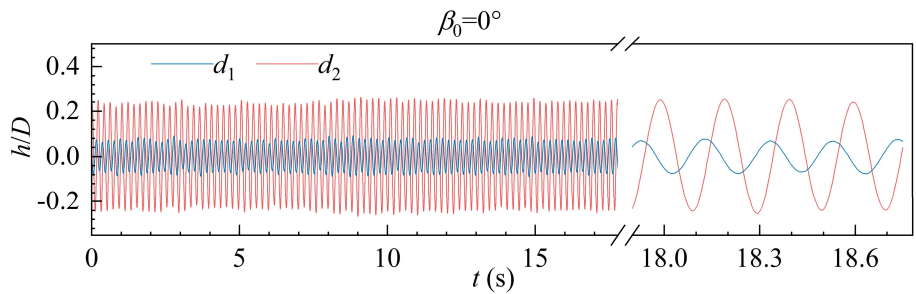

Figure 13. Time histories of displacement signals measured at the suspending arm for the closed-box girder model during conventional coupled VIV occurring at high reduced velocities under normal wind conditions. $d_1, d_2$ are consistent with those in Figure 9.

### 3.3. Heave-torsion coupling eccentricity

To quantify the degree of heave-torsion coupling, we introduce the coupling eccentricity as

$$e = h / \alpha \quad (1)$$

$$h = A_h \sin(\omega t) \quad (2)$$

$$\alpha = A_\alpha \sin(\omega t + \theta) \quad (3)$$

where $h$ and $\alpha$ are the transient heaving and torsional displacements during VIV, respectively. $A_h$ and $A_\alpha$ are the stable heaving and torsional amplitudes. $\omega$ denotes the circular frequency and $\theta$ is the phase difference.

According to Eq. (1), for pure heaving motion or when torsional displacement approaches zero, the calculated eccentricity $e$ tends to infinity. In contrast, during heave-torsion coupled vibration, eccentricity remains finite. Figure 14 illustrates the calculated eccentricities at stable-amplitude stages for torsional VIVs (Figure 10c) and heaving VIV (Figure 11a). In torsional VIV, the eccentricity $e$ fluctuates around a mean of about 0.043, tending toward infinity when the torsional displacement approaches zero. In heaving VIV, the eccentricity *e* oscillates between positive and negative infinity, indicating that the coupling of torsional motion is minimal, which is consistent with the observation in Section 3.2.

Figure 15 shows the evolution of mean eccentricity across various wind yaw angles for torsional VIV, with the mean values and confidence intervals derived from the statistical distribution of the calculated *e* during stable-amplitude stages. The mean eccentricity exhibits a clear increasing trend along with wind yaw angle, suggesting that heave-torsion coupling intensifies as $\beta_0$ increases.

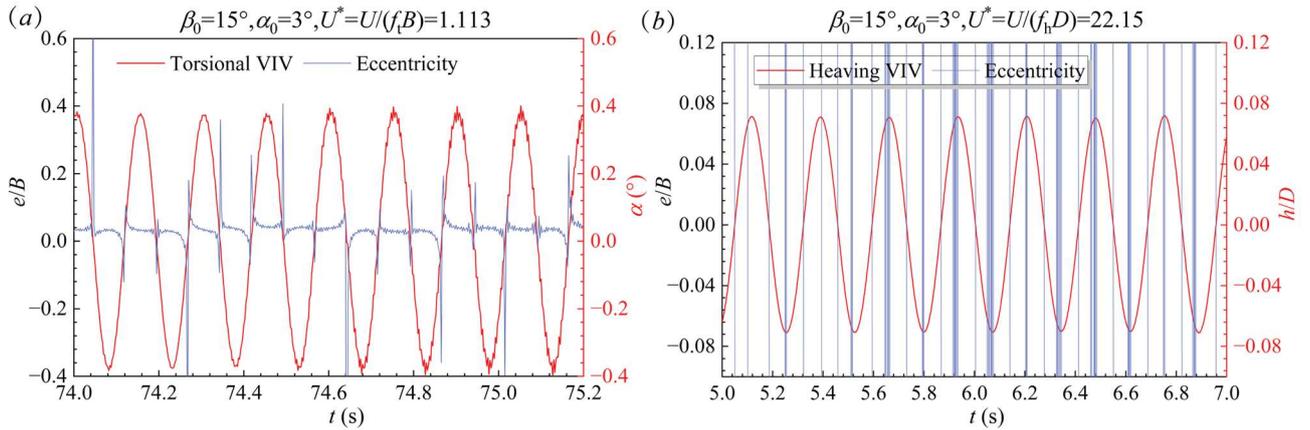

Figure 14. Time histories of vibration eccentricity during stable-amplitude stages for: (a) Torsional VIV in Figure 10c, (b) Heaving VIV in Figure 11a.

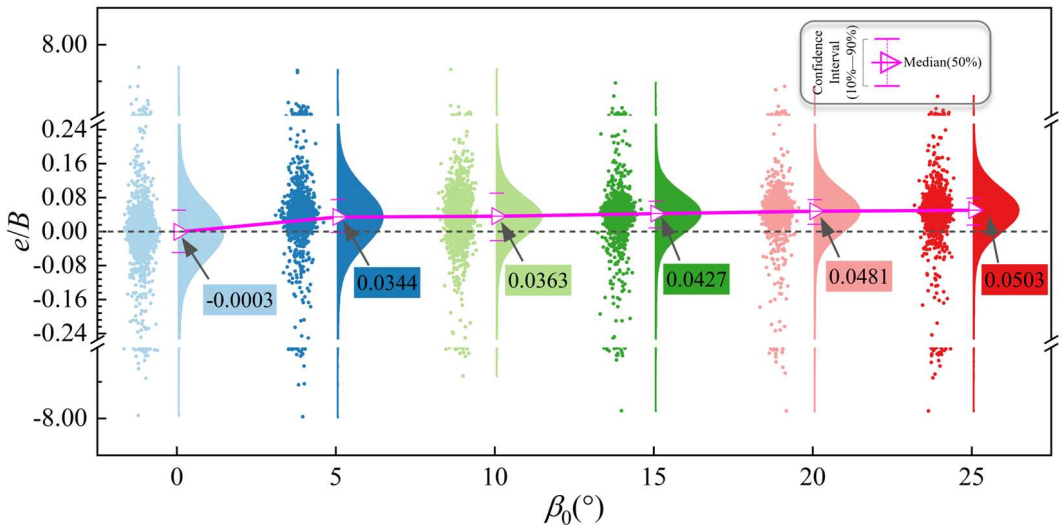

Figure 15. Mean eccentricity at each wind yaw angle for torsional VIV of the closed-box girder model at a 0° wind angle of attack.

## 3.4. Theoretical analysis of heave-torsion coupled VIV under skew winds

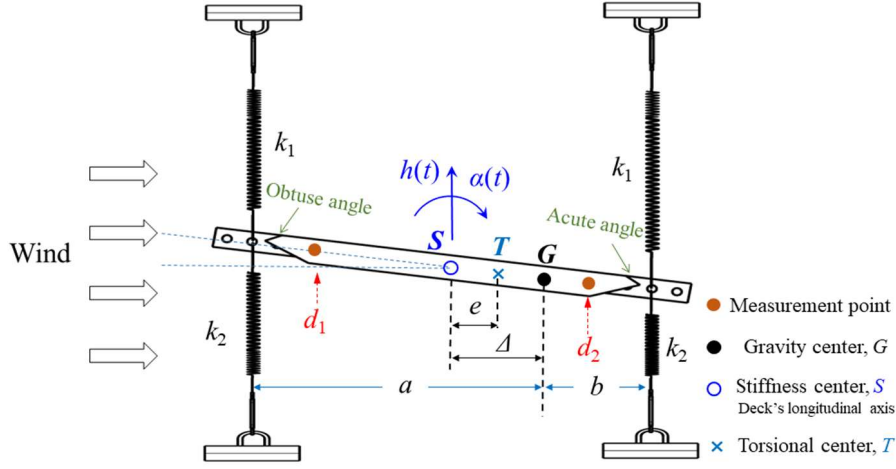

Figure 16. Conceptual diagram of eccentric vibration in oblique section models. $d_1$ and $d_2$ represent displacement signals (positive upward) at the upstream and downstream measurement points, respectively. Point $S$, $T$, $G$ denote the center of stiffness, transient torsional center and center of gravity center, respectively, with $S$ along the bridge deck's longitudinal axis. $h(t)$ and $\alpha(t)$ are the heaving and torsional displacement at the stiffness center $S$. $k_1$ and $k_2$ signify the linear axial stiffness of the upper and lower helical springs, respectively. $e = h/\alpha$ is the transient eccentricity between the torsional and stiffness centers. $\Delta$ is the distance between the center of gravity and stiffness center. $a$, $b$ represent the distance from the center of gravity to the upstream/downstream spring anchor points, respectively.

The observed heave-torsion coupled VIV in oblique section models arises from the asymmetry of the two trapezoidal end segments about the vertical axis. This asymmetry shifts the center of gravity away from the center of stiffness, tilting the torsional axis toward the acute angle sides of the end segments. Figures 2 and 16 depict this shift, with Figure 16 illustrating the view along the section's longitudinal axis. The complex coupled VIV phenomena described in Section 2.2 can be analyzed using the theoretical framework presented below.

Consider an elastically suspended system with effective mass $m$ and moment of inertia $I$ around the gravity center (point $G$ in Figure 16). Let $h$ and $\alpha$ represent the heaving and torsional displacements about the stiffness center (point $S$, along the deck's longitudinal axis). Then the displacements around the center of gravity are derived from the geometric relationship:

$$h^* = h - \alpha\Delta, \quad \alpha^* = \alpha \tag{4}$$

For free vibration in still air, considering only inertia and elastic restoring forces, the Newton's second law applied to oscillations about the center of gravity $G$ yields:

$$\begin{cases} m\ddot{h}^* + 2kh^* + k(a-b)\alpha^* = 0 \\ I\ddot{\alpha}^* + k(a^2+b^2)\alpha^* + k(a-b)h^* = 0 \end{cases} \tag{5}$$

where $k = 2(k_1 + k_2)$ is the effective stiffness of the upstream/downstream helical springs. Substituting Eq.(4) and the geometric relationship:

$$a - b = 2\Delta \tag{6}$$

into Eq.(5) produces the governing equation expressed in terms of the displacements around the stiffness center $S$:

$$\begin{cases} m\ddot{h} - m\Delta\ddot{\alpha} + 2kh = 0 \\ I\ddot{\alpha} + k\left(a^2 + b^2 - 2\Delta^2\right)\alpha + 2\Delta kh = 0 \end{cases} \tag{7}$$

Eq.(7) reveals that the misalignment of gravity center $G$ and elastic center $S$ introduces cross terms involving $-m\Delta\ddot{\alpha}$ in the first equation and $2\Delta kh$ in the second equation, signifying inherent heave-torsion coupling. This coupling persists unless $\Delta = 0$, requiring elastic stiffness symmetry about the center of gravity. However, the asymmetric end segments of oblique section models, aligned with the incidence flow, render heave-torsion coupling unavoidable.

For free vibration, assume harmonic solutions $h = He^{i\omega t}$ and $\alpha = \Theta e^{i\omega t}$. Substituting into Eq.(7) yields

$$\begin{bmatrix} -m\omega^2 + 2k & m\Delta\omega^2 \\ 2\Delta k & -I\omega^2 + k\left(a^2 + b^2 - 2\Delta^2\right) \end{bmatrix} \begin{bmatrix} H \\ \Theta \end{bmatrix} = 0 \tag{8}$$

For non-trivial solutions, the determinant must be zero, resulting in a quadratic equation in $\omega^2$ with two natural frequencies $\omega_1$ and $\omega_2$, for the heaving and torsional modes, respectively. The mode shape of each frequency provides a mode vector with ratio $H/\Theta$, which is generally non-zero, indicating coupled heaving and torsional motion. Consequently, the section model does not exhibit pure torsional or heaving motion but displays heave-torsion coupling. In the absence of damping, the mode shapes are real, implying that heaving and torsional responses are either in phase or out of phase, with a mode-specific fixed torsional center distinct from the center of gravity or stiffness.

In this undamped case, each mode has a fixed instantaneous center of rotation. However, the finite width of trajectories in Figure 12b and the elliptical shape in Figure 12a suggest that the torsional center oscillates during coupled VIV, a phenomenon attributed to perturbations from damping terms, particularly aerodynamic damping. Incorporating damping, the governing equations are expressed in matrix form as:

$$\mathbf{M\ddot{q}} + \mathbf{C\dot{q}} + \mathbf{Kq} = 0 \tag{9}$$

$$\mathbf{M} = \begin{bmatrix} m & -m\Delta \\ 0 & I \end{bmatrix}, \mathbf{C} = \begin{bmatrix} c_{s,h} + c_{a,h} & 0 \\ 0 & c_{s,\alpha} + c_{a,\alpha} \end{bmatrix}, \mathbf{K} = \begin{bmatrix} 2k & 0 \\ 2\Delta k & k\left(a^2 + b^2 - 2\Delta^2\right) \end{bmatrix}, q = \begin{bmatrix} h \\ \alpha \end{bmatrix} \tag{10}$$

where $\mathbf{M}$, $\mathbf{C}$, $\mathbf{K}$ are the mass, damping and stiffness matrixes, respectively. $c_{s,h}$ and $c_{s,\alpha}$ represent the structural damping coefficients in heaving and torsional degree of freedom, respectively. $c_{a,h}$ and $c_{a,\alpha}$ signify the aerodynamic damping coefficients in heaving and torsional degree of freedom, respectively.

Introducing damping terms into Eq.(9) typically perturbs the undamped real modeshapes, making them complex. For instance, during torsional VIV, negative aerodynamic damping destabilizes the torsional degree of freedom, while positive damping stabilizes the heaving degree, resulting in a non-proportional damping matrix. A similar pattern occurs in heaving VIV. However, at low reduced wind velocity, where aerodynamic stiffness is negligible and damping forces are small compared to inertia and elastic restoring forces, the non-proportional damping only slightly perturbs the real mode shapes. In contrast, conventional torsional VIV in Figure 12c exhibits heave-torsion coupling driven by asymmetric aerodynamic stiffness terms, with non-proportional aerodynamic damping becoming significant at high reduced wind speeds [38-39]. Additionally, signature turbulence and vortices induced at the model ends may contribute to stochastic vibrations observed in the phase diagram.

Thus, the observed coupled VIV primarily stems from the mass eccentricity of the trapezoidal end segments. The heave-torsion coupling originates in real modes, with slight complex mode effects from aerodynamic perturbations and stochastic excitation by signature turbulence and vortices, causing the trajectories in Figures 12a and 12b to exhibit finite width rather than straight lines. This mechanism underlies the coupling effects observed during heaving and torsional VIVs in Figures 9-11.

### 3.5. Numerical validation of theoretical framework

In the following section, we propose a preliminary validation of the proposed theoretical framework in Section 3.4.

Since directly measure the mass moment of inertia of oblique end segments is challenging, we employed a numerical approach to determine the mass moment of inertia $I$ and the gravity eccentricity $\Delta$ via Eqs. (7)-(8) based on the heaving and torsional frequencies under skew winds. The procedure is as follows.

(*i*) Determine the spring suspending distance

For a null skew angle, i.e., the symmetric end segments without mass eccentricity, we calculate the spring suspending distance using Eq.(7). By setting the $\Delta = 0$ and substituting into Eq.(7), we obtain:

$$k = \frac{m\omega_h^2}{2}, \quad a + b = 2\sqrt{\frac{I}{m}}\frac{\omega_t}{\omega_h} \tag{11}$$

where $m=ML$ is the total effective mass. $I = J_m L$ is the design value of mass moment of inertia. $M$, $L$ and $J_m$ are listed in Table 1. $\omega_h$, $\omega_t$ are heaving and torsional circular frequencies in still air under null skew angle, respectively.

(*ii*) Identify mass moment of inertia $I$ and gravity eccentricity $\Delta$

For a specific yaw angle $\beta_0 \neq 0$ with oblique end segments, we measure the heaving and torsional frequencies $\omega_{h,*}$, $\omega_{\alpha,*}$ from free decay responses in still air. Due to the asymmetry of oblique end segments, the mass moment of inertia $I$ deviates from the design value. We substitute $\omega_{h,*}$, $\omega_{\alpha,*}$ as eigenvalues into the coupled heave-torsion system in Eq.(8), expressed as

$$\det(\mathbf{K} - \omega^2 \mathbf{M}) = 0 \tag{12}$$

where the stiffness and mass matrix are expressed as

$$\mathbf{K} = \begin{bmatrix} m & -m\Delta \\ 0 & I \end{bmatrix}, \mathbf{M} = \begin{bmatrix} 2k & 0 \\ 2\Delta k & k(a^2 + b^2 - 2\Delta^2) \end{bmatrix} \tag{13}$$

In the above identification of $\Delta$ and $I$, the variation of mass moment of inertia is considered and the geometric relationship of Eq.(6) is employed.

(*iii*) Reconstruct mode shapes and coupling ratio

Using the calculated $\Delta$ and $I$, the mass and stiffness matrices are reconstructed with the calculated parameters. The generalized eigenvalue problem

$$\mathbf{K}\phi = \omega^2 \mathbf{M}\phi \tag{14}$$

is then solved to obtain the eigenvectors $\phi$, which represent the mode shapes of the coupled heave-torsion system. The eigenvalues correspond to the natural frequencies, while the eigenvectors describe the relative contributions of heave and torsion in each mode. In particular, the ratio between the heave and torsion components of each eigenvector is used to quantify the degree of modal coupling, that is the heave-torsion coupling eccentricity $e$ is obtained as

$$e_1 = \phi(1,1)/\phi(2,1), \quad e_2 = \phi(1,2)/\phi(2,2) \tag{15}$$

where $e_1, e_2$ represent the coupling ratios for the heaving and torsional modes, respectively.

Following the above procedure, we conduct modal analysis of the oblique section model system under two typical yaw angles to obtain coupled mode vectors and compare the calculated heave-torsion coupling eccentricity $e$ with the experimental results. The dynamic parameters and calculated results are listed in Table 2. One can find that the mass moment of inertia $I$ slightly differs from the design value under skew wind conditions, which is due to the asymmetry shifts of gravity center away from the stiffness center. The calculated heave-torsion coupling eccentricity $e$ is close to the experimental results.

Fig.17 further illustrates the phase diagrams of heave-torsion coupling VIV using the calculated modal vectors and measured responses. It is found that the calculated modal vectors can satisfyingly reconstruct the observed heave-torsion coupling trajectories during both heaving and torsional VIV. The experimental responses exhibit certain random perturbations around the calculated real modes, which is possibly due to the influence of minor perturbations by damping effect and signature turbulence as discussed in Section 3.4.

Table 2. Dynamic parameters of closed-box girder model and comparison of calculated heave-torsion coupling eccentricity under skew winds during torsional VIV. The total effective mass $m$ is 14.63 kg.

| Skew angle $\beta_0$ | Heaving frequency $f_h$ (Hz) | Torsional frequency $f_t$ (Hz) | Mass inertia moment $I$ (kg·m$^2$) | | Gravity eccentricity $\Delta$ | Heave-torsion coupling eccentricity $e$ | |
|---|---|---|---|---|---|---|---|
| | | | Design value | Calculated | | Measured | Calculated |
| 0° | 4.015 | 6.706 | 0.9310 | | 0 | 0 | 0 |
| 5° | 4.004 | 6.738 | 0.9310 | **0.9270** | 0.0249 | **0.0344** | **0.0387** |
| 20° | 4.001 | 6.745 | 0.9310 | **0.9265** | 0.0281 | **0.0481** | **0.0435** |

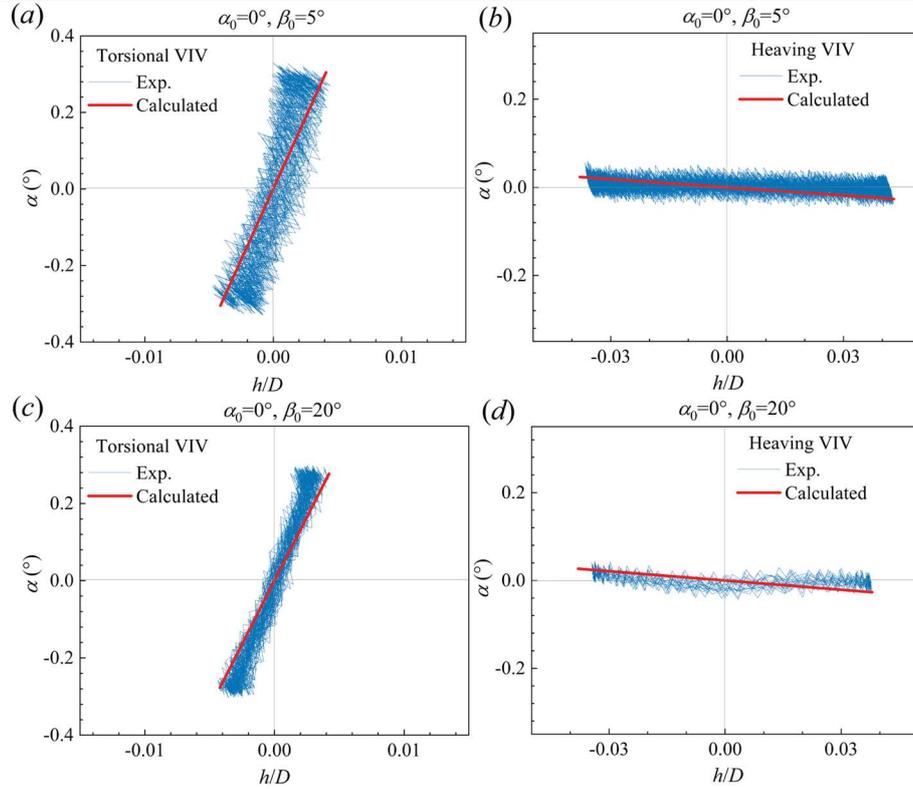

Figure 17. Comparison of phase diagrams during stable stages of heaving and torsional VIVs under skew wind for closed-box girder model.

## 4. Evolution of peak VIV amplitudes with yaw angles

This section examines the revision of peak VIV amplitudes to account for variations in structural damping ratios, slight differences in mass properties due to end segments, and changes in the effective moment of inertia resulting from the torsional axis shift described in Section 3. A numerical method employing the Griffin plot to adjust peak amplitudes is introduced in Section 4.1. The revised peak VIV amplitudes across various wind yaw angles are presented and analyzed in Section 4.2, followed by an assessment of the Independence Principle's (IP) validity.

### 4.1. Correction of VIV amplitudes using Griffin-plots

As noted in Section 3.1, measured VIV amplitudes vary across different wind yaw angles due to fluctuations in structural damping ratios (see Figure 5). Additionally, the replacement of end segments and eccentric coupled vibrations introduce variations in the effective mass properties of spring-suspended oblique models. Consequently, the mass-damping parameters—specifically the Scruton number—differ for each wind yaw angle. Given the sensitivity of VIV responses to these parameters, revising the amplitudes to account for Scruton number variations is essential.

For heaving VIV, Scruton number is defined as

$$Sc = \frac{4\pi m \xi_h}{\rho BD} \tag{16}$$

and for torsional VIV, it is defined as

$$Sc = \frac{4\pi I \xi_t}{\rho B^3 D} \tag{17}$$

where $m$ and $I$ represent the effective mass and mass moment of inertia per unit length of the section model oscillatory system, respectively. Note that $I$ has incorporated the variation due to the asymmetry of oblique end segments using the proposed method in Section 3.5. $\xi_h$ and $\xi_t$ denote the structural damping ratios for the heaving and torsional modes, respectively.

The Griffin plot is utilized to adjust for differences in mass-damping parameters. As shown in Figure 18, the Griffin plot at the peak amplitude state for each wind yaw angle is numerically estimated and used to correct the peak VIV amplitude by interpolating to a unified baseline Scruton number, $\overline{Sc}$.

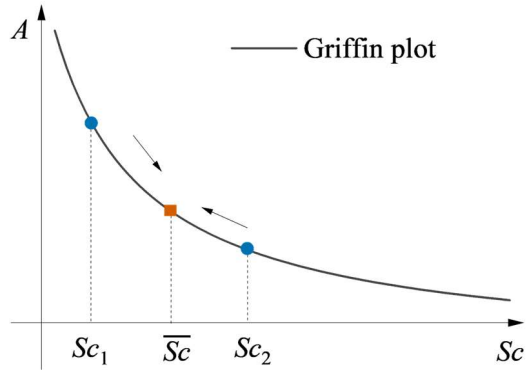

Figure 18 Illustration of compensating peak VIV amplitudes via its Griffin plot.

The numerical procedure is outlined as follows:

(1) For each wind yaw angle $\beta_0$, at the reduced velocity corresponding to maximum amplitude, record multiple grow-to-resonance (GTR) responses from wind tunnel tests.

(2) Extract local peak amplitudes from the GTR responses and fit the transient displacement envelopes using the Richards function (generalized logistic function):

$$A(t) = \frac{A_{max}}{\left[1 + Q e^{-\lambda(t-t_0)}\right]^{1/\nu}} \tag{18}$$

where $A(t)$ is the displacement envelope. $A_{max}$ is the asymptotic maximum amplitude; $\nu$ is the shape parameter controlling the symmetry and steepness of the S-shaped envelope; $\lambda$ is the growth rate; $Q$ is related with initial condition. These parameters are determined through nonlinear least squares fitting.

For comparison, the modified Gompertz function is also employed:

$$A(t) = (A_{\max} - A_0) e^{-b \cdot e^{-\lambda t}} + A_0 \qquad (19)$$

where $A_0$ is the initial amplitude; $b$ is the scaling parameter reflecting the initial damping delay or lag in amplitude growth; $A(t)$, $A_{\max}$ and $\lambda$ are as defined above.

(3) Differentiate the fitted Richards and Gompertz functions to derive the damping curve and then transform it into the Griffin plot.

(4) Utilize the Griffin plot to normalize the peak VIV amplitude to a unified baseline Scruton number. To avoid extrapolation errors, the baseline Scruton number is chosen as the maximum value observed across all tested wind yaw angles for any specific wind angle of attack, ensuring that only interpolation is employed.

This numerical estimation of Griffin plots from GTR displacement responses relies on their self-similarity property; for a detailed analysis and validation, refer to [21].

Figure 19 illustrates the fitting of displacement envelopes and the resulting Griffin plots for the heaving VIV of the closed-box girder model. Both the Richards and modified Gompertz functions effectively fit the displacement envelopes, as depicted in Figure 19a. Figure 19b shows Griffin plots derived from the fitted Richards functions of three GTR responses, demonstrating good agreement and repeatability. In contrast, Griffin plots from the Gompertz function exhibit unphysical distortions at large Scruton numbers, though only the small damping range is relevant for amplitude correction in this study. Thus, both functions are suitable for amplitude interpolation.

Figure 20 compares the corrected VIV amplitudes with the original experimental results, with shaded error regions calculated from 2–4 GTR processes. After applying the Scruton number correction, the amplitude-yaw angle curves exhibit smoother trends than the uncorrected data. The corrected amplitudes are reduced, as the baseline damping ratio corresponds to the maximum value for each wind angle of attack.

The validity of the Independence Principle (IP) is subsequently evaluated by analyzing whether the corrected peak VIV amplitudes remain consistent across wind yaw angles when normalized by the normal component of the flow.

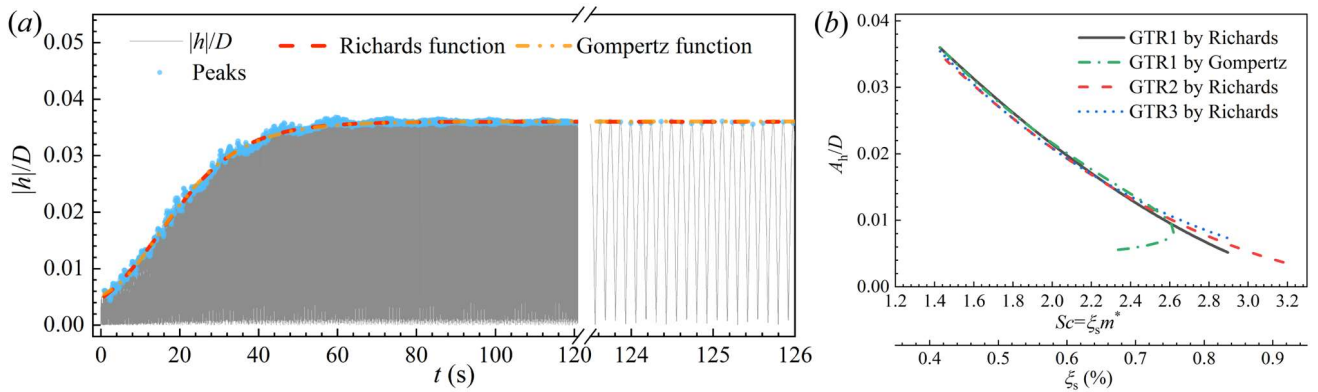

Figure 19. Estimation of Griffin plots from GTR responses of the heaving VIV for the closed box model: (a) the fitting of displacement envelopes with Richards and modified Gompertz functions; (b) the resulting

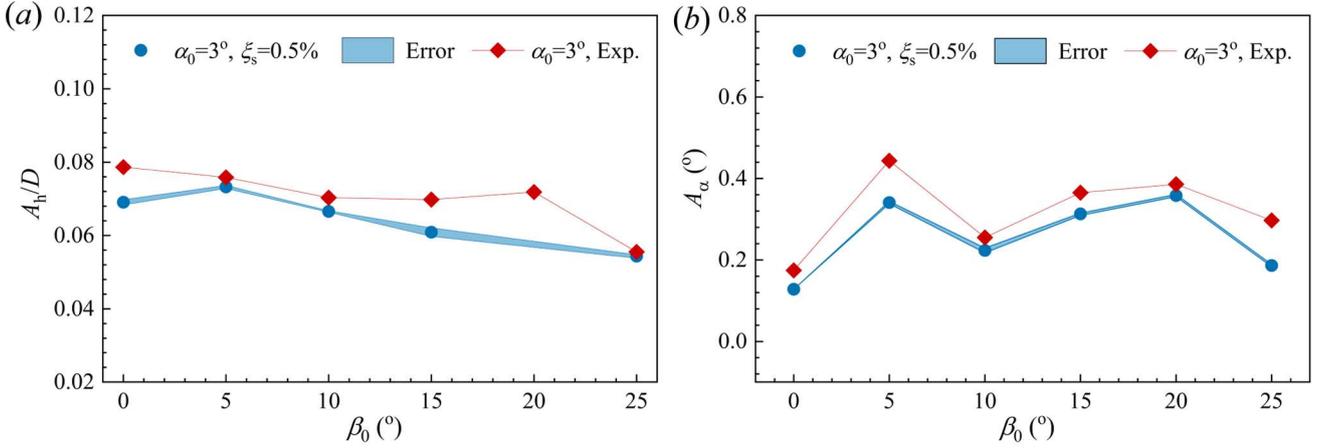

Griffin plots for the heaving VIV.

Figure 20. Comparison of revised amplitude with original experimental results. (a) heaving VIV for twin-edge girder section, (b) heaving VIV for closed box girder section.

### 4.2. Influence of skew winds on peak VIV amplitudes

Using the correction technique introduced in Section 4.1, we revised the VIV amplitudes across all tested wind yaw angles and wind angles of attack. The variation of peak VIV amplitudes with wind yaw angle is of significant engineering importance, as it determines whether conventional experiments under normal flow conditions ($\beta_0 = 0°$) can accurately estimate the maximum VIV amplitude.

For the closed-box girder model, the revised peak amplitude curves are presented in Figure 21. The trends reveal the following:

(1) Both heaving and torsional VIV peak amplitudes exhibit non-monotonic variations with wind yaw angle, consistent with observations on flutter onset wind velocity by Zhu et al. [32-33].

(2) For heaving VIV, the maximum amplitude occurs under normal wind conditions ($\beta_0 = 0°$) at a wind angle of attack $\alpha_0 = 3°$. However, at $\alpha_0 = 0°$ and $-3°$, peak amplitudes are observed at wind yaw angles of 5° and 20°, respectively, with increases of 10.7% and 20.1%.

(3) Torsional VIV is significantly affected by yawed winds, with peak amplitudes occurring at wind yaw angles of 5° and 20°. The amplitude increases by 179.8% at $\alpha_0 = 3°$, 74.1% at $\alpha_0 = 0°$, and 173% at $\alpha_0 = -3°$.

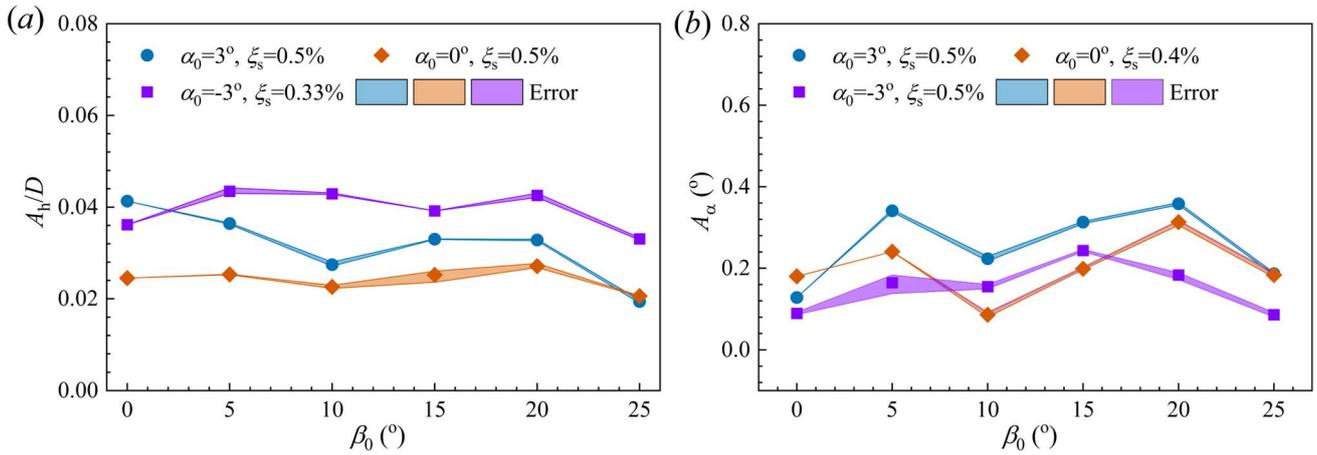

Figure 21. Variation of VIV peak amplitudes with yaw angle $\beta_0$ for the closed-box model: (a) Heaving VIV; (b) Torsional VIV.

For the twin-edge girder model, Figure 22 illustrates that at wind angles of attack of 3° and 0°, peak amplitudes occur at a wind yaw angle of 5°, with increases of 6.1% and 3.9%, respectively. At $\alpha_0$ = -3°, amplitudes remain largely insensitive to variations in wind yaw angle.

The validity of the Independence Principle (IP) is evaluated using Figures 21 and 22. The IP predicts that peak VIV amplitudes should remain approximately constant across different wind yaw angles. However, this does not hold for most tested wind angles of attack, particularly for torsional VIV of the closed-box girder section.

Note that the precise proportions of amplitude changes with respect to wind yaw angle are applicable primarily in experimental contexts, where structural and wind parameters are rigorously controlled. In scenarios involving significant variability in these parameters, such as on-site field conditions, it is necessary to employ appropriate statistical measures, including confidence intervals, to account for amplitude variability.

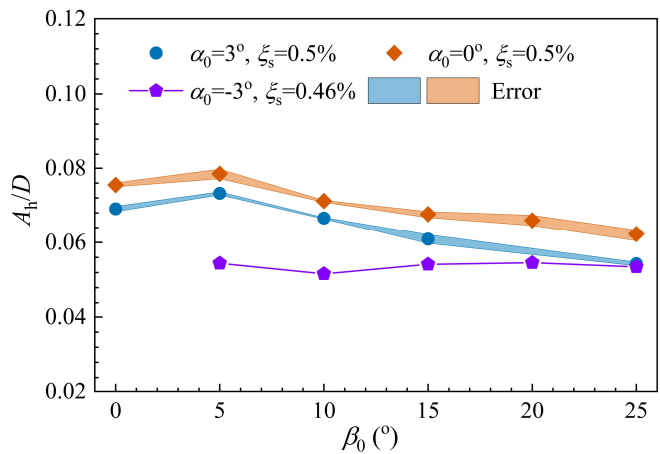

Figure 22. Variation of VIV peak amplitudes with wind yaw angle $\beta_0$ for the twin-edge girder model.

### 4.3. Further discussion

This section addresses additional considerations related to the flow mechanisms, experimental methodologies, and practical implications of the study.

Firstly, it is worthwhile to discuss the possible flow mechanisms underlying the occurrence of peak VIV amplitudes under yawed wind conditions. Based on the existing literature, the amplification mechanisms induced by yawed wind can be categorized into three aspects:

(1) Coupling between axial vortices and Kármán vortex shedding amplifies aerodynamic force fluctuations. Matsumoto et al. [40] observed that axial vortices in the wake of yawed cables interact periodically with Kármán vortices; when their shedding frequencies exhibit 1/3 or 1/5 ratios, the intensity of Kármán vortices is intermittently enhanced, increasing the lift fluctuation amplitude by approximately 40%. Xu et al. [24] further confirmed strengthened vortex shedding regularity around a yaw angle of 20°, with the Strouhal number approximately 0.6% higher than at normal incidence.

(2) Axial flow regulates the wake structure, optimizing the amplitude and phase matching of aerodynamic forces. Direct numerical simulations by Bourguet et al. [25] showed that at yaw angles of 10°–25°, axial flow balances vortex stability and wake diffusion, increasing the fluid excitation energy ratio to 65% (compared to 50% at 0° wind incidence). This delays shear layer separation, maintaining the lift-displacement phase difference within ±15° (optimal for resonance) at a 20° yaw angle. Particle image velocimetry data from Franzini et al. [26] verified quasi-parallel vortex shedding under these conditions, with a lift coherence coefficient of 0.85 (versus 0.7 at 0°).

(3) Enhanced spanwise force correlation. Zhang et al. [36] found that yaw angles of 5°–20° reduce the lift phase difference to ±20° (versus ±45° at 0°), enabling superposition of excitation forces. The spanwise correlation coefficient (0.7–0.9) is substantially higher than at 30° (0.5), therefore avoiding energy dispersion.

These flow mechanisms theoretically support the observed phenomenon; however, existing studies are primarily confined to simple bluff bodies such as circular cylinders, with limited research available on bridge deck-type structures. Future particle image velocimetry experiments are needed to visualize vortex coupling and spanwise coherence at different yaw angles and to quantify their relationships with vibration amplitudes.

While section model tests provide valuable insights into VIV responses and aerodynamic mitigation strategies, their limitations, such as restricted spanwise variations and artificial coupling from oblique end segments, necessitate validation through full-bridge aeroelastic model tests to ensure accurate representation of spanwise aerodynamic force correlations and reliable application to flexible bridge structures.

Considering the significant impact of yawed wind on VIV responses, bridge design practices should prioritize the evaluation of skew wind effects. It is recommended that designers assess VIV amplitudes and the effectiveness of aerodynamic mitigation measures under yawed wind conditions, with a minimum yaw angle range of 0°–25° to capture critical responses.

## 5 Conclusion

The effects of skew winds on vortex-induced vibration (VIV) were examined for two common bridge deck designs—a closed-box girder and a twin-edge girder—through wind tunnel tests using oblique section models. This study analyzed the heave-torsion coupling phenomenon and the impact of wind yaw angles on VIV responses. The primary conclusions are as follows:

(1) Under yawed wind conditions, both heaving and torsional VIV exhibited a heave-torsion coupling effect, manifesting as a single-mode instability. This coupling was significantly stronger in torsional VIV, with the eccentricity of the torsional axis increasing with the wind yaw angle. In contrast, the coupling in heaving VIV was minor and can generally be disregarded.

(2) A theoretical model was formulated to interpret the observed coupled VIV. The heave-torsion coupling stems from the asymmetry of the trapezoidal end segments, causing the center of gravity to deviate from the center of stiffness. The motion predominantly oscillates in a real mode, with minor complex mode effects arising from non-proportional aerodynamic damping.

(3) A numerical algorithm was developed to adjust VIV amplitudes, accounting for variations in structural mass-damping parameters across different wind yaw angles. This algorithm employs the Griffin plot, derived from transient displacement envelopes during VIV, utilizing the self-similarity property of these plots.

(4) The conventional Independence Principle (IP) proved largely inapplicable to both heaving and torsional VIV under skew winds. Both the lock-in range and peak amplitude displayed significant variations with wind yaw angles, challenging the IP's assumption of consistent responses based on normal flow components. The most severe VIV responses for both bridge decks occurred under yawed wind conditions, with peak amplitudes rising by approximately 20.1% for heaving VIV and 179.8% for torsional VIV in the closed-box girder, and by 3.9% for heaving VIV in the twin-edge girder, compared to normal wind conditions.

These findings are based solely on section model tests. Further studies using aeroelastic model tests, numerical simulations, or particle image velocimetry (PIV) measurements are recommended to assess skew wind effects and explore the underlying flow mechanisms. Given the distinct Reynolds number regimes between section model tests and full-scale bridges, large-scale section model tests are essential to minimize scale effects and enhance the applicability of experimental results to full-scale bridge designs.

**CRediT authorship contribution statement**

**Guangzhong Gao**: Writing-Original Draft, Writing-Editing, Conceptualization, Software, Methodology, Formal analysis, Funding Acquisition. **Pengwei Zhang**: Writing-Original Draft, Software, Validation, Software, Methodology. **Wenkai Du**: Writing-Editing, Software, Writing-Review. **Yonghui Xie**: Methodology, Validation, Writing-Review. **Pengjie Ren**: Supervision, Methodology, Writing-Review. **Xiaofeng Xue**: Conceptualization, Methodology, Writing-Review.

**Acknowledgment**

The funding was provided by the National Natural Science Foundation of China, 52278478, by the Research Funds for the Interdisciplinary Projects, CHU (No. 300104240923), Fundamental Research Funds for the Central Universities, CHD (No. 300102214914).

**Declaration of Competing Interest**

The authors declare that they have no competing financial interests of personal relationships that could have appeared to influence the work reported in this paper.

**Data Availability**

Data will be made available on request.